# Nanoscale electronic transparency of wafer-scale hexagonal boron nitride

Caleb Z. Zerger[1,2], Linsey K. Rodenbach[1,3], Yi-Ting Chen[1,2], Ben N. Safvati[1,3], Morgan Brubaker[1,3], Steven Tran[1,3], Tse-An Chen[4], Ming-Yang Li[4], Lain-Jong Li[4,5], David Goldhaber-Gordon[1,3], Hari C. Manoharan[1,3]*

[1]*Stanford Institute for Materials and Energy Sciences, SLAC National Accelerator Laboratory, Menlo Park, CA 94025, USA*

[2]*Department of Applied Physics, Stanford University, Stanford, CA 94305, USA*

[3]*Department of Physics, Stanford University, Stanford, CA 94305, USA*

[4]*Corporate Research, Taiwan Semiconductor Manufacturing Company (TSMC), Hsinchu, Taiwan*

[5]*Department of Mechanical Engineering, University of Hong Kong, Pok Fu Lam, Hong Kong*

*To whom correspondence should be addressed (manoharan@stanford.edu)

Abstract

Monolayer hBN has attracted interest as a potentially weakly interacting 2D insulating layer in heterostructures. Recently, wafer-scale hBN growth on Cu(111) has been demonstrated for semiconductor chip fabrication processes and transistor action. For all these applications, the perturbation on the underlying electronically active layers is critical. For example, while hBN on Cu(111) has been shown to preserve the Cu(111) surface state 2D electron gas, it was previously unknown how this varies over the sample and how it is affected by local electronic corrugation. Here, we demonstrate that the Cu(111) surface state under wafer-scale hBN is robustly homogeneous in energy and spectral weight over nanometer length scales and over atomic terraces. We contrast this with a benchmark spectral feature associated with interaction between BN atoms and the Cu surface, which varies with the Moiré pattern of the hBN/Cu(111) sample and is dependent on atomic registry. This work demonstrates that fragile 2D electron systems and interface states are largely unperturbed by local variations created by the hBN due to atomic-scale interactions with the substrate, thus providing a remarkably transparent window on low-energy electronic structure below the hBN monolayer.

Background and Motivation

Monolayers of hexagonal boron nitride (hBN) have emerged as a valuable building block within the burgeoning materials class of van der Waals (vdW) heterostructures. Applications for hBN vary and include active-layer spacing and isolation[1–3], softening self-gating effects from trapped charge[1,4–7], finessing twist-angle engineering[8–10], and passivation of fragile or ambient-sensitive surfaces[11–13]. These materials and devices can include exfoliated or as-grown hBN. While many such devices explore new ground states involving the emergence of electronic order, magnetism, or superconductivity, there are nearer term efforts in the semiconductor industry to exploit 2D insulators—for which hBN has arisen as the prototype—within conventional silicon-based chip technology[14,15]. In these nanotechnology efforts, CVD growth is preferred for large-area coverage and compatibility with both metallic and insulating substrates. Regardless of the time horizon of research and development, it is critical to understand how the simplest insulator affects the electronic structure of the element it covers in the fabrication process.

Toward this end, we examine one of the cleanest large-area insulator-metal systems: the interface between a monolayer of hBN and a two-dimensional electron system (2DES). The hBN monolayers are grown on Cu(111) using a new wafer-scale growth process developed in the semiconductor industry[16]. Cu(111) on sapphire provides an isolated and natural 2DES which exists as a surface state trapped between a Cu band gap along the [111] direction on the substrate side, and the insulating (vacuum or hBN) interface on the other side. Although hBN can be grown on other metals such as Ir[17,18], Rh[19], Ru[20], Cr[21], Fe[22], Ni[23–25], Pd[26,27], Pt[28,29], we have chosen the Cu(111) system because of its demonstrated compatibility with wafer-scale technology and because its surface state is exceeding well characterized. The 2D electrons within this surface state have been used for a variety of fundamental experiments involving coherent quantum nanostructures and atom manipulation accessing quantum phase[30], quantum spin[31], and designer quantum materials[32]. In principle the as-grown hBN/Cu(111) interface should be a benchmark for electronic homogeneity. Furthermore, hBN has been proposed as a Cu interconnect diffusion barrier to protect neighboring active regions on chip[33], and hence has promise as a passivation or protection layer on complex quantum materials. Since these materials typically must be studied with the hBN in place, its perturbation on the underlying quantum material must be considered.

In this study we have used scanning tunneling microscopy and spectroscopy to gather large spectral maps of the Cu(111) 2DES and neighboring spectral features so that statistical variations in the surface state and neighboring Moiré spectral features can be quantified and spatially indexed to the hBN-Cu(111) Moiré unit supercell.

Results and Discussion

The Moire pattern arising from the lattice mismatch and rotation angle between Cu(111) and hBN is seen in Fig. 1a,c,d, acquired in constant-current mode at $V_s = 4$V and $I_s = 167$ pA. The Moire wavelength shown here is $\lambda = 3.0$ nm, corresponding to a rotation angle $\varphi \approx 4.6°$ assuming a lattice mismatch of 2.2%[16]. This pattern persists over multiple Cu step edges, with the orientation of the Moire pattern and A, B, and C sites preserved as well, consistent with single crystalline character of the hBN monolayer. In Fig. 1b ($V_s = -2.5$ V and $I_s = 33$ pA) we see atomic resolution of the clean hBN monolayer.

We identify three sites of interest as labelled in Fig. 1a inset: the bright spots of the Moire pattern (site A), and sites between the bright spots, one less dark (site B) and one more dark (site C). Each of these sites forms a triangular lattice of the same period, consistent with a pattern arising from different atomic registries. As detailed below, the A sites are associated with an N atom on top of a copper atom and B in an fcc position ($N_{top}B_{fcc}$) as suggested in previous work on morphology of hBN/Cu(111) systems[7,34], appearing as bright spots in the Moire pattern at $V_s = 4$V due to a the CBM being closer to the Fermi level for this registry, leading to a larger local density of states.

The $dI/dV$ features of the hBN/Cu(111) system are dominated by the hBN bandgap as shown in Fig. 2a). The bandgap is ~6 eV but can vary locally by hundreds of meV, in agreement with past STM, ARPES, and spectroscopic measurements[7,35]. At lower junction resistance we can probe features within the gap. For $dI/dV$ spectra taken at a current setpoint $I_s = 167$ pA and a stabilization voltage $V_s = -700$ mV (well within the hBN band gap) we see two consistent features which we highlight in Figs. 2b-c). Figure 2b shows a peak in the differential conductance at near -530 mV. This feature appears on multiple regions of the sample, but is not a feature that appears in $dI/dV$ on a bare Cu(111) surface[36], implying it appears only due to the presence of the hBN layer on Cu(111). To investigate this energy we fit this peak to a Gaussian and find the maximum, and label this energy $E_{BN}$. The other feature highlighted in Fig. 2c shows the characteristic broadened step-edge line shape of the Shockley-type Cu(111) surface state, as observed in other STM work on hBN/Cu(111) systems[7,34]. The intact surface state suggests the interaction between the Cu(111) surface and hBN monolayer is sufficiently weak so that the surface state remains intact, and we are tunneling through the hBN layer directly into the Cu(111) surface state. We fit the midpoint of the rising side of the broadened step function, representing the onset of the surface state, and label this energy $E_{2DES}$.

We investigate the behavior of the $E_{BN}$ and $E_{2DES}$ over multiple terraces of hBN on Cu(111) as shown in Fig. 2d-f. These terraces are separated by over a micron, but each shows the Moire pattern characteristic of monolayer hBN on Cu(111). On each terrace we acquired ~1000 $dI/dV$ spectra over a 5nm x 5nm area, and fit the spectra as described above to find $E_{BN}$ and $E_{2DES}$ for each curve. For each terrace we observe similar behavior for $E_{BN}$: the mean $E_{BN}$ ranges from -519 mV to -534 mV, within one standard deviation of each other, suggesting relative uniformity from one terrace to another. However, within each terrace the distribution of $E_{BN}$ is significantly broader than the mV energy resolution.

In contrast, $E_{2DES}$ has a much tighter distribution within each terrace, with standard deviations smaller than the energy resolution. However, the mean varies from -317 mV in Fig. 2d to -243 mV in Fig. 2f, representing an upshift of 123 to 197 mV. Other STM studies of hBN/Cu(111) systems have observed very different shifts in surface state energy[7,34,37], with the shift sensitive to adsorption energy and local disorder, which vary across the sample at larger length scales.

In Figure 3 we observe the spatial variation in spectral weight of $E_{BN}$ and $E_{2DES}$ for two different terraces. Figs. 3a and d shows the topograph acquired simultaneously with the spectral maps shown in Figs. 3b-c and 3e-f, respectively. These maps are constant energy slices near $E_{BN}$ and $E_{2DES}$ of the spectral maps used in Figs. 2d and f. The topographs were acquired at constant current with setpoint $I$ = 200 pA and $V$ = -700mV.

We observe that for each map the spectral weight of the $E_{BN}$ peak varies much more than the $E_{2DES}$ peak, by nearly a factor of 10. The variation in spectral weight of $E_{2DES}$ is comparable to the variation in spectral weight at other energies (see Supplemental Movies 1 and 2). This suggests hBN is relatively transparent at low energies with transparency locally independent of the hBN morphology. The $E_{BN}$ peak not only varies in spectral weight considerably more across the map, but the variation is closely correlated with the Moire pattern, with the $E_{BN}$ peak much higher near the A sites of the Moire pattern.

The peak energy $E_{BN}$ correlates with the Moire pattern as well, in addition to the intensity of the peak. Fig. 4c shows the $dI/dV$ spectra from the map shown in Fig. 3d ordered by the location of $E_{BN}$ and color coded by distance to the Moire A sites. One can see that higher $E_{BN}$ (closer to the Fermi level) is correlated with spectra acquired closer to the Moire A sites. The opposite is true for the B sites, and no significant correlation is found between $E_{BN}$ and the C sites (see Supplemental Figure 1).

In striking contrast, the location of $E_{2DES}$ (Fig. 4d) does not correlate with distance to the Moire A site (nor with B or C). In addition to being uncorrelated with the Moire pattern, the total range of $E_{2DES}$ is close to the error bars of the measurements, set by resolution and confidence intervals of the fits (see Figure 4d). Notably the variation is nearly an order of magnitude smaller than the range of ~100 meV for $E_{BN}$, and the standard deviation is just 4.1 mV. This reinforces the previous conclusion that, despite electronic structure arising from the hBN/Cu(111) system, at low energies we are able to tunnel through the hBN and observe the surface state relatively unperturbed by local morphology.

From these results we can conclude that the $E_{BN}$ peak arises from interactions between the hBN and Cu(111) associated with the registry of hBN atoms on the Cu(111) surface. These interactions are suggested as explanations of other properties that have been shown to vary with the Moire period such as work function and band gap[34]. Computational work suggests different atomic registries may produce a range of different adsorption energies and interaction energies varying on a scale of tens of meV per unit cell[38], a similar scale to the variation we see in $E_{BN}$ over the Moire pattern. The work function is also known to vary between high symmetry registries, with calculated values of work function variation of 90-130 meV[37,38] and measured values of work function modulation of 60 meV for a Moire wavelength of 3.5 nm, again comparable to the shift we see in $E_{BN}$ over the Moire pattern. The shift in work function is suggested to arise from the hybridization of B and N atoms with Cu atoms.

We deduce that $E_{BN}$ is also related to states arising from B or N atoms interacting with the Cu surface. Previous DFT work[37] has suggested non-zero PDOS near -500 meV arising from the $p_z$ orbitals of the B and N atoms, with a shift in energy of ~60 to 100 mV depending on the atomic registry. The correlation of the peak shift (toward the Fermi energy) with the A site and anti-correlation with the B site suggests these are locations of opposite extrema in the location of these interaction states, which is consistent with an $N_{top}B_{fcc}$ registry for the A site and $N_{fcc}B_{hcp}$ registry for the B site (the C site would then be $N_{hcp}B_{top}$). Because $E_{2DES}$ does not vary significantly with the Moire pattern we can conclude that the surface state is largely unaffected by these local interactions between the B and N atoms with the Cu(111) surface.

However, $E_{2DES}$ does vary greatly between terraces, suggesting a large variation of adsorption energy across the sample. The upshift in $E_{2DES}$ varies from 123 to 197 mV on the terraces studied, corresponding to adsorption energies of 69.2 to 111.0 meV/unit cell, following the prescription suggested by Ziroff et. al.[37,39]. The variation in adsorption energy may be due to local strain[40], defects or disorder over certain terraces, or a stacking fault in the underlying Cu substrate, which are minimized but have finite probability from the growth parameters[16].

Conclusions

In summary, we have investigated two spectral features of a hBN/Cu(111) system, one associated with the interaction between the two surfaces ($E_{BN}$), and another the native Cu(111) surface state $E_{2DES}$. We observe a large variation in $E_{BN}$ correlating with the Moire pattern, which we associate with high symmetry points of the atomic registry and corresponding hybridization between hBN and Cu atoms.. Despite these interactions the surface state remains remarkably locally homogeneous, showing no variation on the scale of the Moire wavelength within experimental error. This result is encouraging for the prospect of hBN as a weakly interacting layer in van der Waals heterostructures, and for applications targeting isolation of fragile underlayers without perturbing low-energy electronic structure.

Author Information

CZZ and HCM designed experiments. CZZ acquired STM/STS data. YTC, BNS, MB, and LKR performed data analysis. ST and LKR prepared samples. TAC performed hBN/Cu sample growth. MYL characterized the hBN using Raman spectroscopy. LJL supervised the growth and characterizations. DGG coordinated collaboration and advised the project. CZZ, LKR, and HCM wrote the manuscript with input from all authors. HCM supervised the project.

Acknowledgements

This work is supported by the U.S. Department of Energy, Office of Science, Basic Energy Sciences, Materials Sciences and Engineering Division, under Contract DE-AC02-76SF00515. hBN samples were grown at the TSMC Corporate Research Smart Lab in Hsinchu, Taiwan.

**Methods:**

**Growth of Cu (111) thin films.** The as-received sapphire substrates were first etched in a mixed $H_2SO_4/H_3PO_4$ aqueous solution at 300 °C for 20 min. The etched sapphire substrates were cleaned by immersion in ultra-pure water for 5 min. After cleaning, the sapphire substrates were loaded into a sputtering chamber for Cu deposition. The chamber was maintained at room temperature under an argon pressure of 0.3 mTorr, giving rise to a Cu deposition rate of 2 nm s$^{-1}$.

**Chemical vapour deposition of *h*BN.** The monolayer *h*BN films were grown in a three-inch furnace tube with three heating zones using low-pressure chemical vapour deposition (LPCVD). The two-inch Cu (111)/sapphire substrate was placed in the central heating zone of the main chamber. Ammonia borane (97%, roughly 60 mg) was used as the precursor and loaded into a subchamber at the upstream side of the main chamber. The furnace was first pumped down to a base pressure of 5.0 Torr. Before growth, the substrate was annealed at 1,050 °C for 60 min under a hydrogen gas flow of 300 standard cubic centimetres per minute (sccm). The subchamber (with precursors) was heated to 85 °C using a heating belt and maintained there for 30 min. The precursor was then introduced into the main chamber for 30 min in order to grow *h*BN on the Cu (111)/sapphire substrates, where the substrates were facing downwards. After *h*BN growth, the subchamber was closed, and the main chamber was naturally cooled to room temperature under a hydrogen gas flow of 30 sccm.

**Sample Preparation.** To prevent oxidation of the Cu(111) layer, the as-grown sample was shipped in a vacuum sealed bag and then stored in an Ar-filled glovebox. The sample was briefly removed from the glovebox so it could be cleaved into a size suitable for mounting in our system (side lengths ≤ 10 mm). To protect the hBN surface during the cleaving process, a LatticeGear FlipScribe 100 was used to scribe the sapphire substrate while the sample remained hBN-side up. LatticeGear cleaving pliers were then used to cleave the sample along the scribe line; care was taken to ensure that the pliers only touched the outermost edge of the sample. After cleaving, the sample was submerged in acetone for 10 minutes, then removed and submerged in methanol for 10 minutes. The sample was then mounted on a sample mount with top contacts and put under ultrahigh vacuum (<$10^{-10}$ Torr) and heated to 540ºC for 24 hours, then at 300ºC for another 12 hours, and finally cooled down to room temperature before transferring the sample to the low temperature STM.

**Measurements.** Measurements were performed at 77 K in a low-temperature UHV STM using a Pt-Ir tip checked for cleanliness and spectral flatness on Au(111). Topographies were taken in constant current-mode, while *dI/dV* spectra were taken using a standard lock-in technique, with typical modulation voltages between 1.4 and 7 mV at a frequency of 586 Hz. Spectral maps were acquired in scans up to 24 hours and typically 961 individual spectra, with a typical spatial resolution of 1.6Å and an energy step of 6.3 meV. Topographs acquired simultaneously with the spectral maps have spatial resolution of 0.14 Å, with topographical data acquired at points between those at which spectral data was acquired.

# Figures

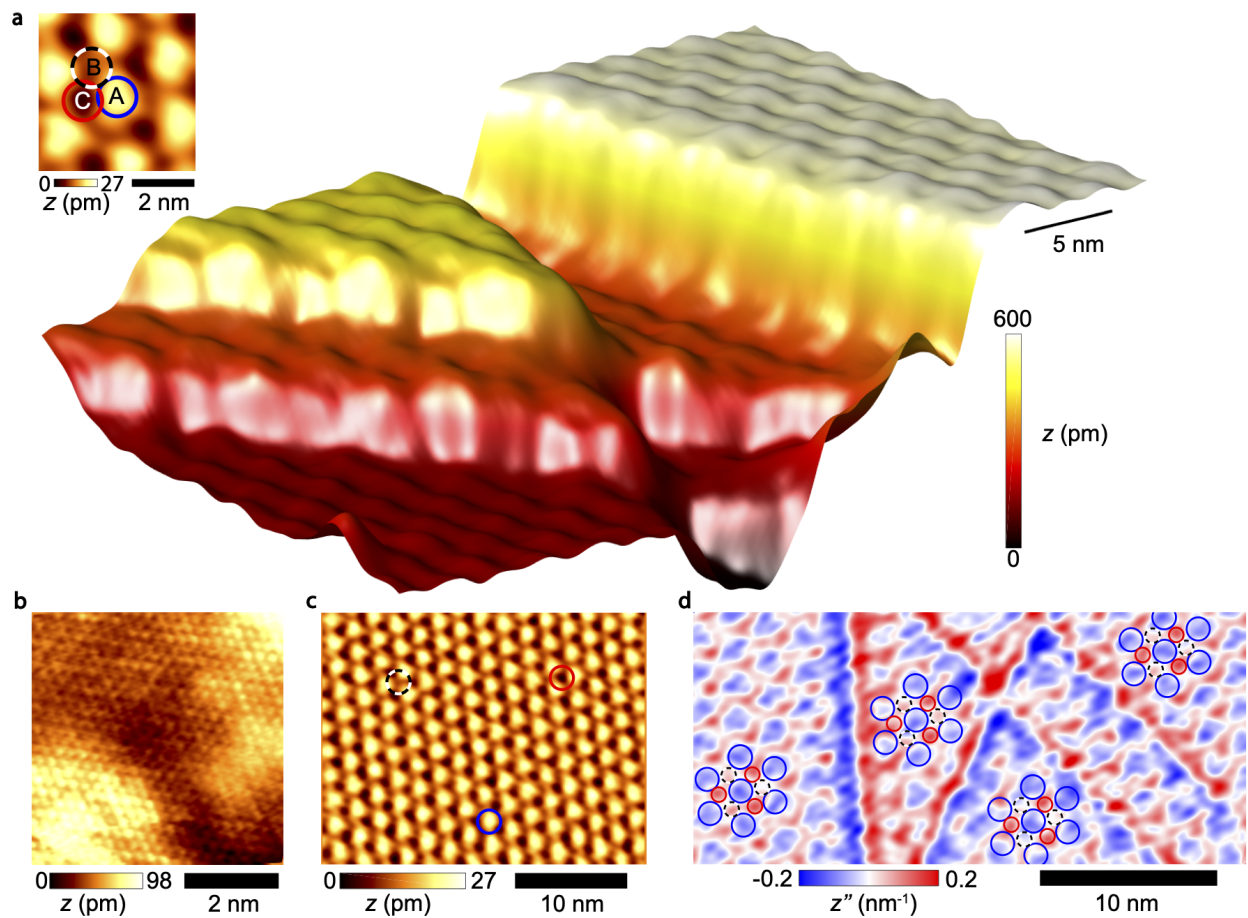

**Figure 1 | Nanoscale views of wafer-scale hBN on Cu(111).** (a) A 3D view of a region of wafer-scale hBN spanning several adjacent atomic terraces on Cu(111) acquired at $V_s$ = 4 V and $I_s$ = 167 pA). Identical Moiré patterns are observed on each step. Inset: a zoomed-in topographic map showing a few unit cells of the Moiré pattern. The three high-symmetry points of the Moiré pattern are circled and labeled. (b) Atomic-scale resolution topograph of the hBN monolayer ($V_s$ = -2.5 V and $I_s$ = 33 pA). (c) Topograph ($V_s$ = 4 V and $I_s$ = 167 pA) of a flat region in a typical terrace displaying a Moiré pattern. The three high-symmetry points of the Moiré pattern are again circled. (d) Curvature map of the region displayed in (a), in which a Moiré pattern is observed on all terraces. A subset of the three high-symmetry sites are circled and replicated over all terraces without scaling or rotation. The coherence of the overlay over multiple terraces is consistent with mono-orientation and the single-crystal nature of the hBN monolayer[16].

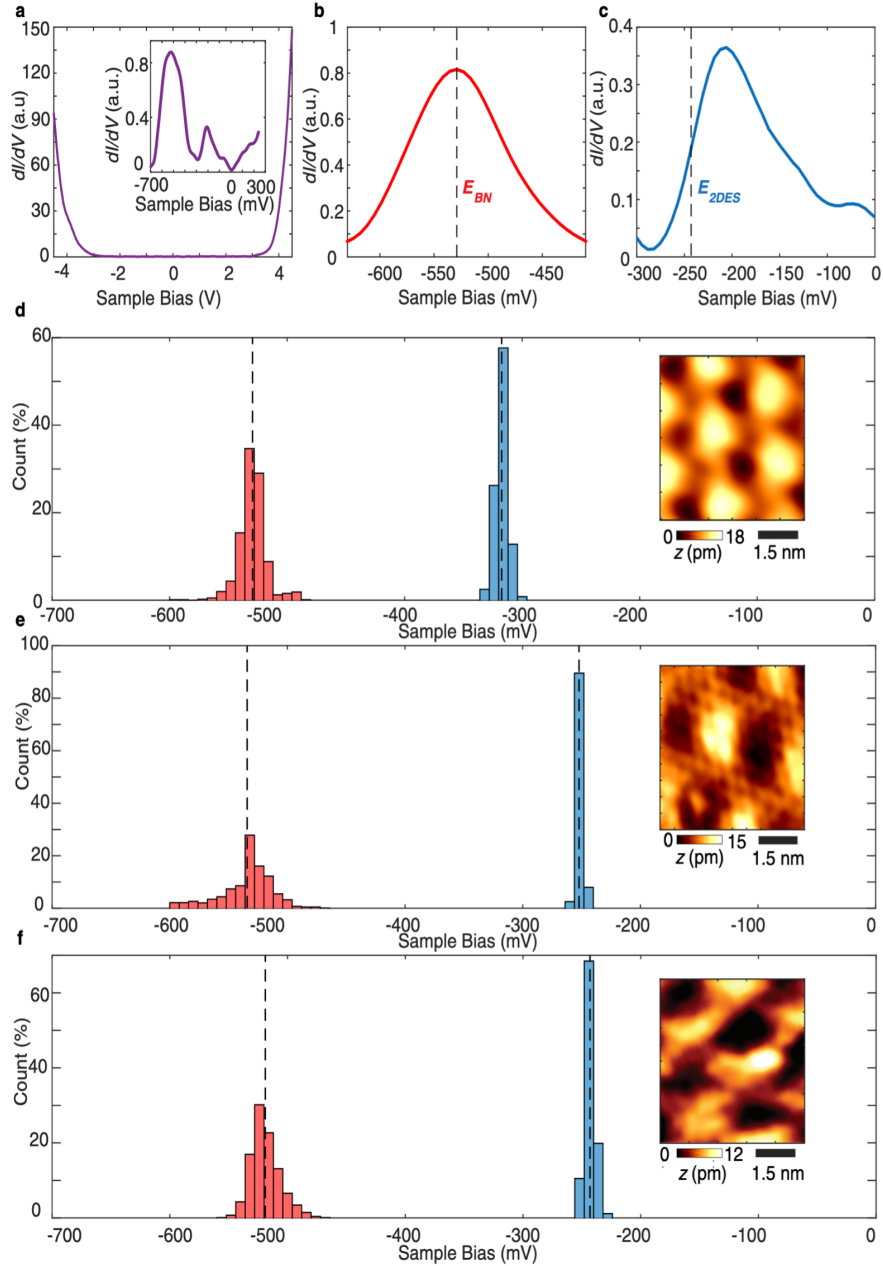

**Figure 2 | Low-energy in-gap tunneling spectra and statistics.** a) *dI/dV* taken with $V_S$ = 4.5 V, $I_S$ =273 pA showing a typical hBN gap of 6.1 V. The inset shows a *dI/dV* spectrum taken at $V_S$ = -700 mV, $I_S$ =167 pA inside the hBN gap, showing two important features, a peak near -530 meV and another near -240 meV. b) -530 meV feature taken from an average over hundreds of nearby *dI/dV* curves (same conditions as (a)). This peak tends to be nearly symmetric and is approximated as a Gaussian for fitting purposes, with the peak energy labelled $E_{BN}$ and c) -240 meV feature also taken from an average of *dI/dV* curves showing the broadened step edge shape associated with the Cu(111) surface state, with the associated energy $E_{2DES}$. d-f) Histograms of $E_{2DES}$ and $E_{BN}$ fit to 961 *dI/dV* spectra from three 5nm x 5nm atomic terraces, showing a broader distribution of $E_{BN}$ within each terrace compared to the variation of $E_{2DES}$. Insets show topographs taken at d) $V_S$ = 4V, $I_S$ =167 pA and e,f) $V_S$ = -700 mV, $I_S$ =167 pA showing the Moire pattern over each terrace.

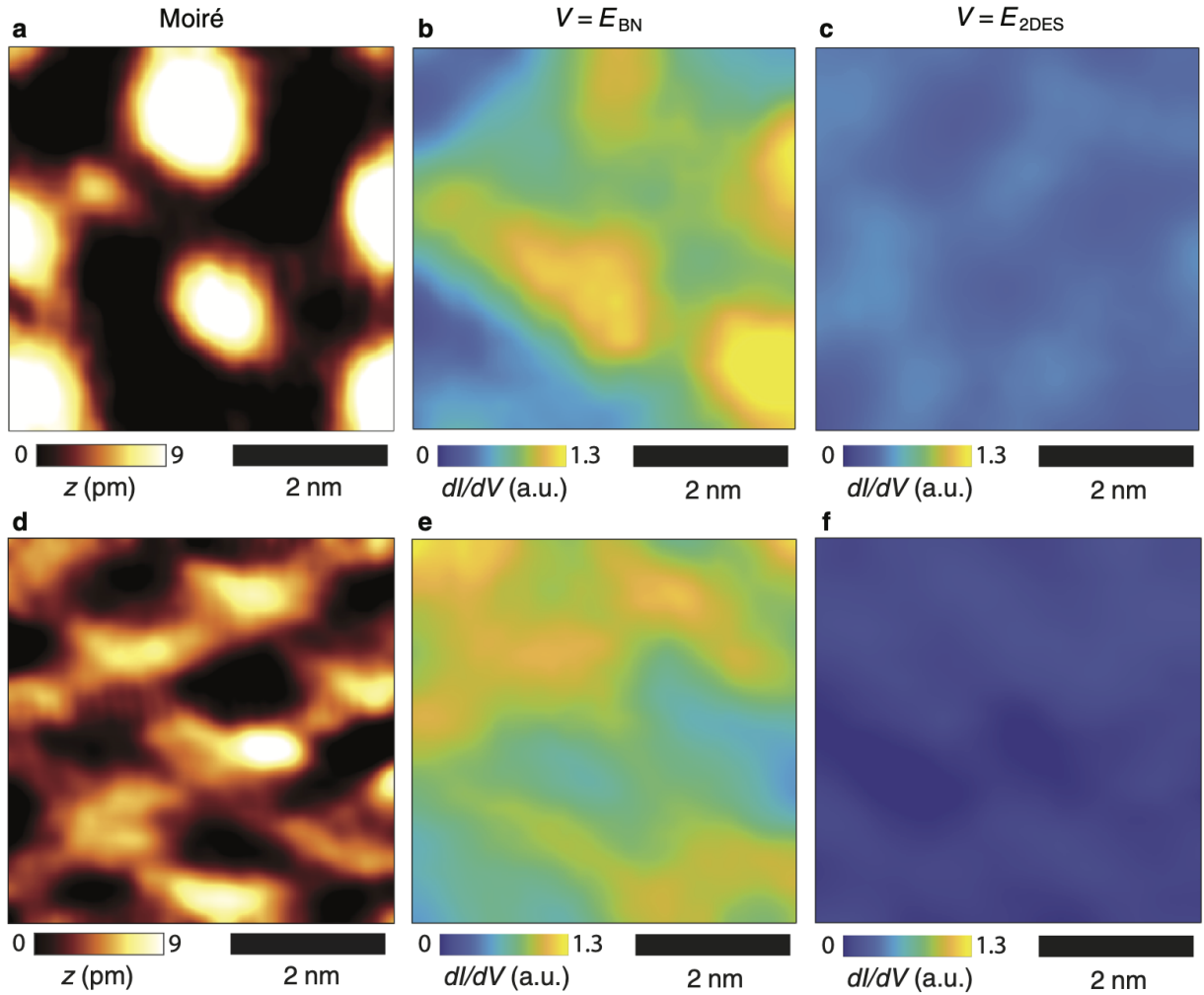

**Figure 3 | Spectral maps over distant Moiré regions.** a, d, Images of the moiré pattern from constant-current topographs on hBN/Cu(111) ($V_s$ = -700 mV and $I_s$ = 167 pA). b,e, Color plot of the differential conductance spectra at the mean $E_{BN}$ energy over all sites on each terrace. Spectra were taken between -700 mV and 250 mV with an energy resolution of 6.3 mV. c,f, Conductance spectra color plot at the mean $E_{2DES}$ onset energy over all sites on each terrace. The evident spatial correlation between dI/dV($E_{BN}$) to the moiré pattern, and the lack of correlation between and the dI/dV($E_{2DES}$) with the moiré, suggests different physical origins of these two electronic features.

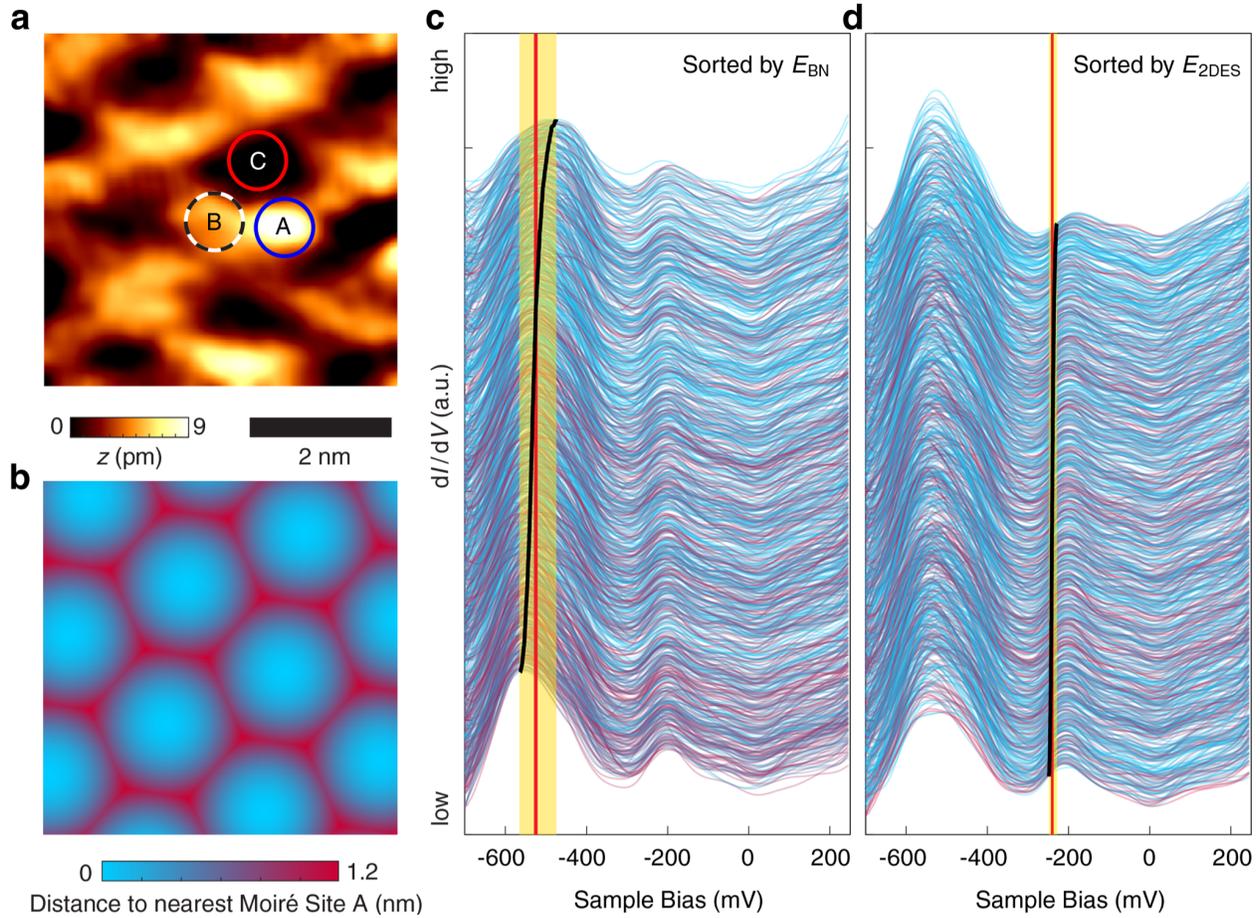

**Figure 4 | Correlation between Moiré pattern and BN and 2DES electronic structure.**
a, STM topograph ($V_s$ = -2.5 mV and $I_s$ = 33 pA) of the Moiré pattern. Red, black/white, and blue circles mark the location of Moiré site A, site B, and site C, respectively. b, Distance to the nearest Moiré site A with the identical field of view as a. c, Waterfall plot showing STM $dI/dV$ spectra acquired simultaneously with the topograph in a. A small offset has been added to each point spectrum to visualize the trends, and the spectra are sorted according to the peak location of $E_{BN}$. The black curve marks the peak location of the spectra. The width of the yellow box is the range of observed peak energy, showing a 90.1 mV range for $E_{BN}$. The width of the vertical red line denotes the error bar of the peak energy. The color of each spectral trace indicates the distance between the location the spectrum was acquired and the nearest Moiré site A. The color scale is the same as b. d, Waterfall plot created using similar methods to c, but here sorted by $E_{2DES}$.